\documentclass{egpubl}
\usepackage{eurovis2022}

\ConferenceSubmission   % uncomment for Conference submission
\usepackage[T1]{fontenc}
\usepackage{dfadobe}  

\usepackage{cite}  % comment out for biblatex with backend=biber
\BibtexOrBiblatex
\electronicVersion
\PrintedOrElectronic
\ifpdf \usepackage[pdftex]{graphicx} \pdfcompresslevel=9
\else \usepackage[dvips]{graphicx} \fi

\usepackage{egweblnk}
\usepackage{booktabs}
\usepackage{xspace}

\newcommand{\fidyll}{\textit{Fidyll}\xspace}
\newcommand{\idyll}{\textit{Idyll}\xspace}

\usepackage{amssymb}
\usepackage{multirow}
\usepackage{pifont}% http://ctan.org/pkg/pifont

\usepackage{xcolor}

\title[\fidyll: A Compiler for Cross-Format Data Stories \& Explorable Explanations]%
      {\fidyll: A Compiler for Cross-Format\\Data Stories \& Explorable Explanations}

\author[M. Conlen \& J. Heer]
{\parbox{\textwidth}{\centering M. Conlen\thanks{Work done at the University of Washington}$^1$\orcid{0000-0002-4992-4066}
        and J. Heer$^2$\orcid{0000-0002-6175-1655} 
        }
        \\
{\parbox{\textwidth}{\centering $^1$Our World in Data\\
    $^2$University of Washington\\
       }
}
}
\begin{document}

% uncomment for using teaser
\teaser{
 \includegraphics[width=\linewidth]{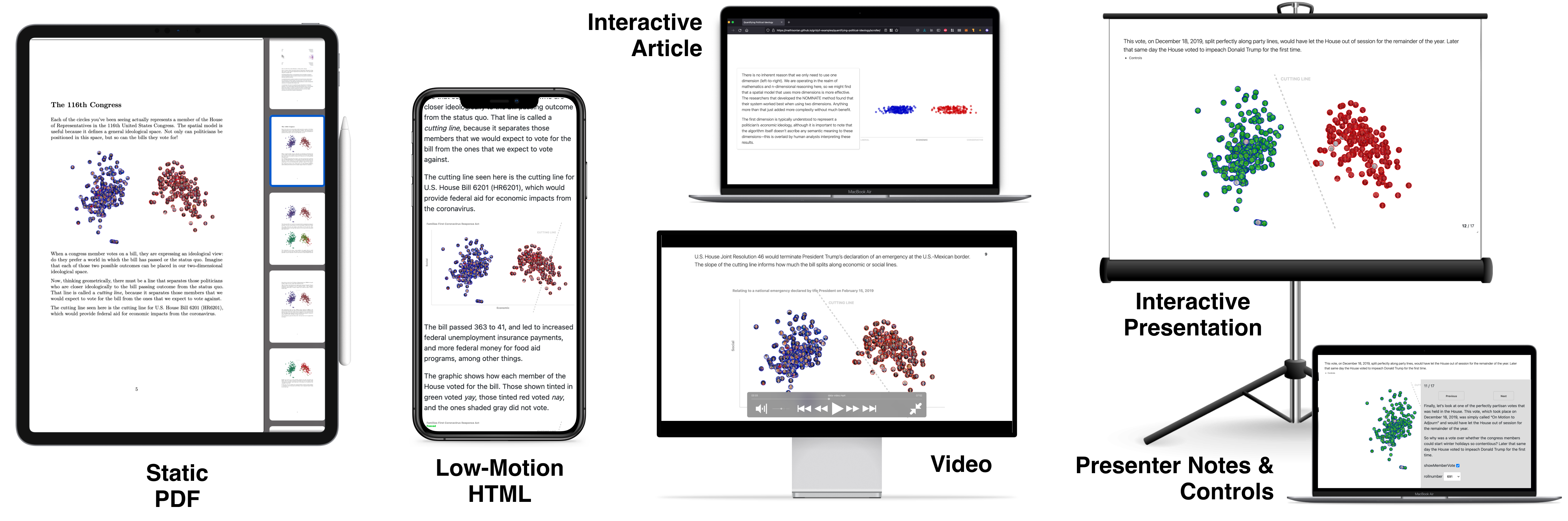}
 \centering
  \caption{\label{fig:teaser}
           \fidyll~supports serializing five different output formats from a single source document. Here different versions of the case study \textit{Quantifying Political Ideology} are shown. The different formats each have their own strengths and weaknesses.}
}

\maketitle

%-------------------------------------------------------------------------
\begin{abstract}
Narrative visualization is a powerful communicative tool that can take on various formats such as interactive articles, slideshows, and data videos. These formats each have their strengths and weaknesses, but existing authoring tools only support one output target. We conducted a series of formative interviews with seven domain experts to understand needs and practices around cross-format data stories, and developed \fidyll, a cross-format compiler for authoring interactive data stories and explorable explanations. Our open-source tool can be used to rapidly create formats including static articles, low-motion articles, interactive articles, slideshows, and videos. We evaluate our system through a series of real-world usage scenarios, showing how it benefits authors in the domains of data journalism, scientific publishing, and nonprofit advocacy. 
% We used the system to develop a narrative visualization, Quantifying Political Ideology, published as a static and interactive article and presentation. 
We show how \fidyll provides expressive leverage by reducing the amount of non-narrative markup that authors need to write by 80-90\% compared to \idyll, an existing markup language for authoring interactive articles. 
% This allows authors to focus on their domain specific content. 

%-------------------------------------------------------------------------
%  ACM CCS 1998
%  (see https://www.acm.org/publications/computing-classification-system/1998)
% \begin{classification} % according to https://www.acm.org/publications/computing-classification-system/1998
% \CCScat{Computer Graphics}{I.3.3}{Picture/Image Generation}{Line and curve generation}
% \end{classification}
%-------------------------------------------------------------------------
%  ACM CCS 2012
%   (see https://www.acm.org/publications/class-2012)
%The tool at \url{http://dl.acm.org/ccs.cfm} can be used to generate
% CCS codes.
%Example:

% \ccsdesc[300]{Computing methodologies~Collision detection}
% \ccsdesc[300]{Hardware~Sensors and actuators}
% \ccsdesc[100]{Hardware~PCB design and layout}

\printccsdesc   
\end{abstract}  
%-------------------------------------------------------------------------
\section{Introduction}
\label{sec:introduction}

Interactive articles~\cite{hohman2020communicating} are a powerful medium for conveying complex, data-driven stories 
to wide audiences in engaging and understandable ways. The format is used by 
major news outlets like the New York Times  (e.g.~\cite{parshina20203}), non-profits and other advocacy groups  ~\cite{rall2016data}, 
and scientific communicators~\cite{dragicevic2019increasing,olah2017research}. These visual narratives can utilize dynamic techniques like personalization~\cite{adar2017persalog} and self-reflection~\cite{kim2017explaining} to increase audience engagement~\cite{2019-idyll-analytics} and improve learning outcomes~\cite{mayer2002multimedia}. However, they tend to be time consuming and difficult to produce~\cite{hohman2020communicating}. 

Moreover, there are other formats (or \textit{genres}) of narrative visualization~\cite{segel2010narrative}---such as slide shows \cite{mckenna2017visual}, lectures~\cite{rosling2001hans}, data videos~\cite{amini2015understanding}, and comics~\cite{bach2017emerging}---each of which may be more appropriate for use with certain audiences or contexts. For example, while a rich interactive narrative like \textit{Snowfall}~\cite{branch2012snow} may be engaging to some audience members~\cite{mckenna2017visual}, others may find the extensive use of animation to be distracting or even disorienting ~\cite{frederick2015effects}. A researcher who publishes a new finding as an interactive article may wish to deliver the same results in the form of an interactive presentation (e.g.,~\cite{conlen2018beginner}); a data journalist who publishes an interactive data story online may need to create a static version of that same article for publication in a print newspaper or in animated format to be shared on social media (e.g.,~\cite{conlen2019launching}). 

While it is clear that there is no ``one-size-fits-all'' format for narrative visualization, authoring tools don't take this into account: existing tools like Ellipsis~\cite{satyanarayan2014authoring} and \idyll~\cite{conlen2018idyll} focus on producing a single artifact---whether it be an interactive article or slideshow, or an annotated visualization---and offer limited support for authors to produce alternative versions. In this work, we explore an authoring tool that facilitates the re-targeting of interactive, data-driven content across multiple formats. With \fidyll, authors write their data story in a high-level markup language and the system compiles this into a standard data schema that can then be used to produce a number of different formats including interactive articles, interactive slideshows, videos, static PDFs, and low-motion web pages. Our contributions include:

\begin{table*}\centering
\footnotesize
\begin{tabular}{@{}l cc p{7cm} p{4.5cm}@{}}\toprule Format & Interaction & Animation & Benefits 
% & Examples 
& References \\ 
\midrule
Scroller & \checkmark & \checkmark & Highly polished designs may elicit more engagement from readers.
% & \textit{Snowfall}~\cite{TK} 
& Gruessing \& Boomgaarden\cite{greussing2019simply}; Conlen et al.~\cite{2019-idyll-analytics}
\\
Stepper  & \checkmark & \checkmark & Supports both synchronous (live presentation) and asynchronous (reading slides as web page) communication. & McKenna et al.~\cite{mckenna2017visual}; Zhi et al.~\cite{zhi2019linking}
\\
Low-motion & \checkmark & x & Graphics are distributed over space instead of time. May support interactivity. Limits motino sickness. & Frederick et al.~\cite{frederick2015effects}
\\
Video & x & \checkmark & Well-supported format with powerful distribution platforms like YouTube. & Amini et al.~\cite{amini2015understanding}; Bradbury \& Guadagno ~\cite{bradbury2020documentary}
\\
Scientific Paper & x & x & Archivable, standardized format & Tversky et al.~\cite{tversky2002animation}
\\
\bottomrule\end{tabular}\caption{\label{table:formats}Interactive document formats like scrollers and steppers are useful because they can support both animation and user interaction, but other formats for narrative visualization may be more appropriate depending on the context and audience.}\end{table*}

\begin{itemize}
    \item We conduct semi-structured interviews with seven domain experts to understand the requirements and workflows associated with multi-format interactive articles. We synthesize findings through an open coding process and use these results to construct motivating usage scenarios. 
\item We build \fidyll, a cross-format compiler for interactive data stories and explorable explanations. \fidyll produces interactive articles, slideshows, PDFs, and videos from a single input source. \fidyll is open source and available for use at ~\url{https://github.com/idyll-lang/fidyll}.
    \item We produce three articles, each of which corresponds to a motivating scenario. Through these case studies we show that \fidyll provides a large amount of expressive leverage by allowing authors to produce multi-format articles while reducing the overall amount of non-narrative markup by up to 90\%. 
\end{itemize}

%-------------------------------------------------------------------------

\section{Background}
\label{sec:bg}

Our work builds directly on research in narrative visualization and interactive visualization authoring tools. We draw inspiration from past work on adaptive layout and re-targeting of content across different display sizes and modalities. Our work is also related to accessibility, archiving, and personalization in narrative visualization. 

\subsection{Narrative Visualization \& Interactive Articles}

Segel \& Heer~\cite{segel2010narrative} identified a set of seven genres of narrative visualization (magazine style, annotated chart, partitioned poster,
flow chart, comic strip, slide show, and film/video/animation). The set of genres has since been expanded to include additional techniques like interactive articles which use scroll-based triggers~\cite{mckenna2017visual}.  Segel \& Heer note that the genres ``vary primarily in terms of (a) the number of frames and  [...]  (b) the ordering of their visual elements.'' We leverage this insight in order to support targeting multiple narrative visualization genres as output from a single input source. Victor described \textit{explorable explanations}~\cite{victor2011explorable} as a technique to combine text and interactive graphics to promote active reading behaviors~\cite{active-reading}.

Interactive articles~\cite{hohman2020communicating} are useful because they support animation and interactivity, allowing authors to take advantage of multimedia learning techniques~\cite{mayer2002multimedia}. However, they are not preferable in all situations or contexts (see Table~\ref{table:formats}). Researchers have found that individual preferences play a key role in audience engagement. For example, McKenna et al.~\cite{mckenna2017visual} found that some readers prefer slideshows to scroll-based articles and engaged more if articles were presented in the format of their preference. Interactivity and animation also present challenges in accessibility: some readers of websites which utilize scroll-based interactivity and parallax will experience motion sickness ~\cite{frederick2015effects}; interactive graphics require additional development to support screen readers and may not be beneficial for visually impaired users~\cite{kim2021accessible}. Additionally interactive articles present a challenge for preservation and archiving; while preservation tools~\cite{kelly2012warcreate} and formats~\cite{mohr2008warc} for rich interactive web content exist, they are seldom incorporated as part of the publishing process~\cite{broussard2018saving} and are not as well supported as static document formats like PDF~\cite{gomes2011survey}.

Tools like Ellipsis~\cite{satyanarayan2014authoring}, \idyll~\cite{conlen2018idyll, conlen2021idyll}, and VizFlow~\cite{sultanum2021leveraging} were created to aid authors in the narrative visualization production process. Ellipsis and VizFlow are relatively high-level but target a limited range of outputs;
% : Ellipsis supports annotated interactive visualizations, and VizFlow supports various scrollytelling layouts but does not support interactivity. 
\idyll is more expressive but requires more markup to achieve similar results; the relationship between \fidyll and \idyll is akin to that of Vega-Lite~\cite{satyanarayan2016vega} and Vega~\cite{satyanarayan2015reactive}, with  \fidyll serializing much of the low-level \idyll code necessary to implement common design patterns. Like Ellipsis, \fidyll utilizes a scene-based data model; however, our model generalizes to support an arbitrary number of parameterized graphics embedded within a larger narrative structure.
% while Ellipsis targets a single annotated visualization. 
In contrast to VizFlow, our tool supports output targets across multiple formats, and \textit{interactive} graphics, which can be manipulated by readers beyond scroll input. 
Because data stories tend to written by collaborators of varying technical proficiency~\cite{lee2015more}, simple markup languages like ArchieML~\cite{strickland_tse_ericson_giratikanon_2015} have been adopted by many of the major newsrooms~\cite{archiemlaward}. ArchieML can be embedded in existing collaborative word processors like Google Docs. Our system utilizes a novel markup language based on ArchieML and of similar syntactical complexity.

\subsection{Adaptive Layout of User Generated Content}

Given the prevalence of mobile devices~\cite{2019-idyll-analytics}, the question of how to adapt content for various display sizes and layouts is of major importance. Hoffswell et al.~\cite{hoffswell2020techniques} described techniques for flexible responsive visualization design, allowing authors to produce visualizations that gracefully adapt between mobile, table, and desktop screens. In this work we treat graphics as parameterizable black-boxes, with the assumption that they can gracefully respond to variously sized displays. 
% This assumption works well in many cases but in practice some authors may need to design their graphics using a system like that described by Hoffswell. 
More generally, the problem of adaptive document layout has been of great interest to researchers in the HCI community~\cite{hurst2009review}. Jacobs et al.~\cite{jacobs2004adaptive} developed a method of grid-based document layout building on the fundamental visual grid-system from the Swiss school of design~\cite{muller1981grid}. In this work we also utilize grid-based templates, but don't attempt to automate fine-grained placement of components. Instead we allow authors to modify the placement of text and associated graphics in terms of the grid when necessary.
% , and instead rely on a set of predefined templates which can flexibly support a range of content. 
In the future an automated adaptive layout algorithm could be incorporated. 
% into \fidyll.

%-------------------------------------------------------------------------

\section{Formative Interviews}
\label{sec:interviews}

To understand the needs and practices of interactive article authors we conducted semi-structured interviews with seven domain experts, including a data visualization practitioner and educator (P1), a data journalist and academic researcher (P2), an artist and scientific communicator who uses data visualization and physicalization as their medium (P3), a machine learning and human-computer interaction researcher (P4), a data journalist and human-centered artificial intelligence researcher (P5), a librarian specializing in physical and digital maps (P6), and a machine-learning researcher (P7). 

\subsection{Methods}

To source the interview participants we directly emailed three people who were known to be domain experts. We also posted a call for participants in a Slack channel dedicated to \textit{explorable explanations}~\cite{explorable}, and on Twitter, where the authors are followed by a number of data journalists and visualization researchers. The interviews took place remotely over a video call, and lasted about thirty minutes. Our interview script (available in the supplementary material) served as a general guideline for the interviews, but we engaged in open discussion with the interviewees, guiding the conversations based on their interests and expertise to better understand their needs. After the interviews were completed we used an open coding procedure to analyze the interviews and identify the most salient themes, presented below. 

\subsection{Qualitative Results}

\textbf{Format is determined by intrinsic and extrinsic factors.} Respondents discussed a  variety of formats that they had used to present narrative data visualizations, including web articles, slide shows, Power Point presentations, Google Docs, PDFs, data physicalizations, magazine articles, print newspaper stories, and interactive maps. The choices for which format to use were driven by a variety of factors, including format affordances (``How do we educate and teach people the capabilities and limitations of machine learning? One way to do that is through interactive articles, through play, education, and other explorable-type interfaces'' P4), audience preferences (``I try to do things as simple as possible... especially for journalistic audiences it's not in your best interest to make a very complicated data visualization'', P2), author expertise \& resources (``We are all in different backgrounds, but we aren't scientists. We don't have a certain workflow to work with.'' P3), author goals \& preferences (``There are some things that research papers don't encourage... I don't know why reviewers don't like empty space, no matter how pedagogical or good the diagram is. These restrictions are good at times, but if you want to write for an audience that wants to learn I feel slideshows or what \textit{Distill} does... those things are nice for reaching a wider audience.'' P7), and externally imposed constraints (``At [national publisher] I would make charts for the web and then have to put them in the magazine... It sometimes went the other way too where they had a feature magazine article, some illustrator would have a graphic that works in print, and then we'd make an interactive version of it.'' P5). 
  
\textbf{Visualizations from exploratory tools are often repurposed for use in a narrative context.} While many authors used JavaScript to create custom web-based interactive visualizations ( ``The tools are your standard web-development tools'' P4), it was also common for authors to create graphics using exploratory analytics tools like R and Python (``We mostly use R and R Studio. But it depends on the story.'' P3, ``We do the interactive version where we don't know what we're looking for and its very exploratory and then we make an explanatory version'' P5.). These graphics may be included directly in articles without additional modification or exported in a vector format for additional manipulation and annotation ( ``I tend to use R for most of the things that I do and will polish in Adobe Photoshop or Illustrator if needed'' P2; ``The assets, the coupling between them, you have some representation of your interactive thing, you save that out once, and then you work with it in another program: you have your SVG crowbar, pull it out, open it up in Illustrator and hope it mostly works.'' P5). In some cases this approach makes it difficult to add interactivity post-hoc (P2, P7).

% . I don't need the version with four charts linked together, if we found the pattern, once you found it you can just make a chart showing the pattern.

 \textbf{It is common to create static and interactive versions of the same content.} Most interviewees had experience adapting materials from an interactive resource to a static one (``For [project] it was visualizing principal component analysis. One of them is in JavaScript on the client, another one is in Python in Deepnote, and then its a PNG in the final report. Three different renderings of the same---it's a huge pain.'' P1). This was done for a variety of reasons, including academic publishing requirements
%  ``Our PDFs are solely online, graphics and all.'' P2; 
(``The dissertation format required a PDF at the end of the day.'' P4), to create an long-living archive of the interactive resource (``Not knowing how long its going to last for, if they're going to have access to the software they used to create it, if they have to present in other formats like a paper, or if they want to go out into the community and work with people who don't have access to digital tools and methodology, how do we show it as a PDF?''), or to create short executive summaries of the main points of their interactive article (P1 ``We want the raw data and a data scientist might, but I don't want to hand you that. What I want to hand you is maybe an interactive report you can scroll through, but the mayor doesn't want that. The mayor wants the last thing which is `here are a few bullet points, who cares what the methodology is, what's the takeaway?''').

 \textbf{Graphics are reused across formats with little modification, but may be omitted from static materials; text may be rewritten to better suit different audiences.} When authors needed to adapt graphics for different formats, they typically did so without making significant changes to the underlying graphic (``There's no special treatment or augmentation [when translating graphics from a static to an interactive version]'' P4). When changes were made, it was typically to add annotations (``We'll take screenshots of the graphics, paste them in slides, and write some annotations on top of it'' P5), or to capture some interactive aspect of a graphic, for example by recording a video or taking a screenshot (``Authoring interactive graphics, primarily for the web but secondarily for a keynote or slide deck... same annoying problem of how do I capture the example, but at least its not a PDF so you can take screenshots or GIFs or videos to show the interactions more in depth'' P4, ``To preserve the interactivity you can record a video and add a transcript to show off what the tool once did and provide contextual information about why you made the choices that you did.'' P6). Interactive graphics may be replaced by small multiples in a static environment (``there were only two states [in the interactive graphic] and you can capture it in a screenshot and show them side-by-side'' P4), or they may be omitted altogether due to the lack of support for adapting to the new environment (``we did a web-only version that was a Leaflet map that people could click on; we couldn't figure out a way to make that easily navigable in print'' P4). Authors frequently changed the text of their articles to better fit the expected audience of that particular format (``Delivering a narrative summary based on the data science results remains a big problem. How do you go from displaying a bunch of raw data and visualizations to constructing a narrative at different levels of abstraction and expertise.'' P1).

 \textbf{Authors want to produce more cross-platform content but may not have the capacity to do so.} Most of the authors had ambitions to create more cross-format content (``It's very important for our organization to be able to deliver narrative summaries of data science in an automated way... right now there are four assets that we produce.'' P1). They noted that they understood how different formats are beneficial for different audiences and contexts (``The interaction should facilitate a faster understanding or learning compared to reading a 10 page PDF. How do we give someone an interactive summary to come away with most of the important stuff? The details we can leave somewhere else...'' P4). However, participants noted that due to time, resource, and format constraints they often don't do this in practice (``We are not technical people. We are not programmers'' P3; ``It's work that is not even seen. If you look at it, I don't think that it matters to you what colors I used or how the elements are arranged.'' P7; P5 ``We're making a nice interactive interface for exploring GAN outputs. It works well.. I can imagine a nice blog post about it, but I'm unsure about how to boil it down to a research paper... it seems so much worse than the blog post version in terms of reading. Why would anyone read this PDF if we could make the version with inline [interactive] examples?'').

 \textbf{Existing tools don't promote accessibility or archivability of interactive content.} While many interactive article authors were focused mainly on what they could do to improve the effectiveness of their publications work (``It should be very precise... the data representation should be accurate. If there's an interaction it should only help the person engage with the results, come away with a technical insight, or learn something new.'' P4), others were concerned about the longevity of these materials however they were lacking resources and guidance. (``We'll have a student who spent a semester creating a really snazzy data project... and they're realising that its hard to preserve that type of work as a portfolio piece... This is the type of question that I get. I have been approaching things on a case-by-case basis as there aren't great best-practices established, its really thinking about the goals of the particular researcher and their situation.'' P6) or how they work for users who may be vision impaired or prefer reduced motion.
 
%  A goal is `can someone view this chart and understand very precisely differences in the underlying data?' 

\section{Motivating Scenarios}
\label{sec:motivating-scenarios}

Based on our interviews with domain experts, as well as the authors' own experience working on data journalism and scientific communication projects, we developed a set of three motivating scenarios grounded in current real-world practices. While not exhaustive, these scenarios are meant to represent the breadth of domains and tasks which demand that authors of interactive narrative data visualizations need to produce content that lives in multiple formats. In \S\ref{sec:evaluation:case-studies} we use \fidyll~to implement a cross-platform article corresponding to each of the motivating scenarios.

\textbf{MS1. Data Journalism}.  
A data journalist wants to use a statistical model as an aide in explaining the voting records of members of the United States House of Representatives. She works with a data scientist to identify various parameters of the model that would make a compelling story. The journalist then writes text that guides readers through these various parameters. The journalist works with a graphics editor to create a visualization that presents the output of the model. The story needs to work on desktop, mobile, and tablet devices, and related static renditions of assets need to be created for a print version of the news paper as well as for use on social media and the paper's homepage. 

\textbf{MS2. Non-profit Advocacy}. 
A watchdog organization regularly publishes reports online discussing the efficacy of various climate change mitigation approaches. These reports are highly technical, consisting of charts, tables, references, and occasionally interactivity. The writers are experts in climate science, policy, and data analysis but they are not experts in web development or graphic design. The organization wishes that all of their stakeholders can participate in the creation of these articles, but currently all of the work falls on to the shoulders of one team member, who happens to have some web development experience. They wish to more easily incorporate the results of statistical models written in R and Python without writing JavaScript and HTML. 
% The nonprofit organization wishes to develop a set of core components, which can then be readily invoked in a parameterized fashion by any of their team members. They wish to be able to develop articles in a collaborative manner, for example through google docs.

\textbf{MS3. Scientific Publishing}. 
A researcher has identified methods to make a kernel density estimation algorithm significantly faster, but with certain trade-offs in the fidelity of the algorithm’s output. The researcher has submitted this work to be published at an academic conference, and needs to write a paper including static images of the output. The researcher will also need to create a slideshow with which to give a presentation, which will be delivered via a pre-recorded video. The slideshow will include animated versions of the graphics used in the paper. In addition, the researcher wishes to publish a digital blog post covering the same topic so that their work can be seen by a wider audience.

\section{\fidyll}

\fidyll is a cross-genre compiler for data stories and explorable explanations. Authors create a single specification which is then used to generate multiple different output formats including interactive articles, dynamic slideshows, data videos, and static documents.

\subsection{Data Model}

\begin{figure}
  \centering
  % the following command controls the width of the embedded PS file
  % (relative to the width of the current column)
  \includegraphics[width=\linewidth]{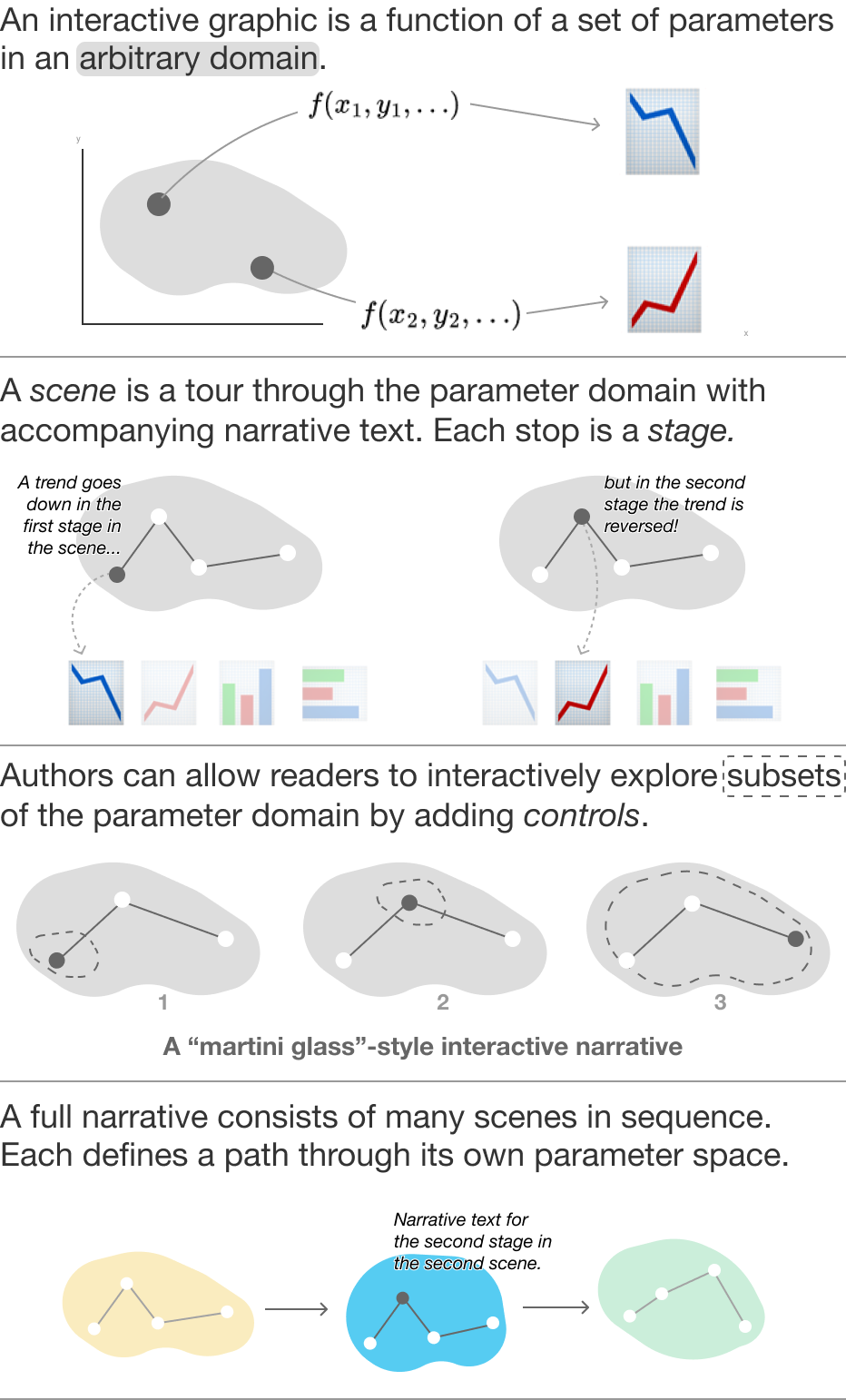}
  % replacing the above command with the one below will explicitly set
  % the bounding box of the PS figure to the rectangle (xl,yl),(xh,yh).
  % It will also prevent LaTeX from reading the PS file to determine
  % the bounding box (i.e., it will speed up the compilation process)
  % \includegraphics[width=.95\linewidth, bb=39 696 126 756]{sampleFig}
  %
%   \parbox[t]{.9\columnwidth}{\relax
%           For all figures please keep in mind that you \textbf{must not}
%           use images with transparent background! 
%           }
  %
  \caption{\label{fig:data-model}
           \fidyll's data model centers on the notion of giving readers a tour through a high-dimensional parameter space. Each \textit{scene} defines its own parameter space, and each \textit{stage} within the scene defines a set of parameter values. \textit{Controls} give readers leeway to explore parts of this space that aren't directly visited by a stage.}
\end{figure}

A crucial piece of \fidyll is the data model (Figure~\ref{fig:data-model}) used to represent interactive narrative visualizations. By leveraging a expressive and flexible schema, all of the constituent narrative pieces can be specified, independent from their arrangement and layout on the page. 

\textbf{Narrative.} The top-level element is the \textit{narrative}. The narrative consists of document metadata such as authors, title, subtitle, etc., along with document-wide information such as pointers to datasets. A narrative consists of a series of \textit{scenes}, which can optionally be book-ended by an introduction and a conclusion. 

\textbf{Scenes.} A \textit{scene} is the primary constituent of \fidyll~data-stories. Each scene consists of a reference to a parameterizable data graphic, along with a list of stages. Scenes may include additional configuration information such as flags to include or exclude particular scenes from particular output genres, allowing authors to customize content on a per-genre basis. 

\textbf{Stages.} A \textit{stage} represents a particular configuration of the data graphic associated with a scene. Each stage specifies a particular configuration of domain-specific parameters for the data graphic, along with corresponding narrative text. Parameters can either be stationary---in which case they take on a specific value as soon as the stage comes into view---or animated, in which case they can loop through a series of values on a timer, or interpolate between two values at the start and end of the stage. Similar to scenes, authors can also use \textit{filters} to include or exclude particular stages from particular output genres. Stages can include a set of \textit{controls}, which allow readers to interact with the data graphic.

\textbf{Controls.} A \textit{control} allows readers to interactively adjust the domain-specific parameters that drive the data graphics embedded in each scene. To add a control, authors specify which parameter it corresponds to, along with the domain of allowed values which readers should be able to specify. \fidyll will then infer which type of widget is appropriate and add it to the document. 

\textbf{Animations.} Authors can define \textit{animations} for each parameter at each stage by providing a \textit{start} and \textit{end} value, and configuration options (e.g. \textit{duration} and \textit{loopcount}); if the start value is not provided the previous value of the parameter is used. In formats that don't support animation, a series of static frames are generated; authors can define the number and arrangement of static frames with the \textit{frames} (number of frames to generate) and \textit{columns} (the number of columns in the resulting grid) options.

\textbf{Filters.} Authors can provide filters for any scene or stage to specify in which formats the contents will appear. There are four filter keywords that are supported: \textit{include}, \textit{exclude}, \textit{only} and \textit{skip}. The \textit{include} and \textit{exclude} options each expect a list of formats, to specify that a stage or scene will only be included in those formats (\textit{include}) or that it will not be included in those formats (\textit{exclude}). The \textit{only} and \textit{skip} keywords are intended for use during article development, and are boolean flags which allow an author to specify that only one particular scene or stage should be rendered (\textit{only}) or that a particular scene or stage should not be rendered during development (\textit{skip}); we include these based on experience developing long interactive articles, it is often necessary to use such functionality to focus on developing a particular subset of the article.

\subsubsection{Output Targets}

The primary tasks of \fidyll~are (1) to parse the author-provided markup into a machine-readable JSON data structure, and then (2) for each of the desired output targets, to transform this normalized data into an \idyll abstract syntax tree corresponding to the respective format.  \fidyll supports five output formats: \textit{interactive article}, \textit{low-motion article}, \textit{PDF}, \textit{interactive slideshow}, and \textit{video}. 

\textbf{Scroller.} The interactive article target implements the popular \textit{scrollytelling} layout and reflows to support viewing across display sizes such as on mobile, tablet, and desktop devices. Each \textit{scene} is displayed with the data graphic fixed to the screen while the text associated with each stage scrolls over top. As each section of text appears on the screen, the parameters associated with that scene are applied to the graphic, which updates its display in response. If a \textit{stage} has \textit{controls} associated with it, then the appropriate widgets are displayed beneath the text and readers can manipulate them.

\textbf{Low-motion HTML.} The static article is laid out as a single column of text interspersed with data graphics. For each stage in a scene, the text is displayed with the data graphic rendered with the relevant parameters. Interactive controls are still included, but animations are converted to a series of still frames. 
% This format may be prefered by users on mobile devices or those who prefer articles that limit the amount 

\textbf{Static Paper.} The static article is laid out in one or two columns of text interspersed with data graphics. No widgets for controls are displayed; instead, an appendix is constructed which renders the relevant data graphic with each of the possible configurations of parameters defined by the cross product of controls. To avoid a combinatorial explosion in the where there are many controls, authors can define a subset of configurations which are to be included in this appendix. Animations are displayed as a series of static frames. 

\textbf{Interactive Slideshow.} The interactive slideshow implements the popular \textit{stepper} layout and can be used in an asynchronous manner (where the audience reads through the slideshow at their own pace), or a synchronous manner where a presenter uses the slides as support during a live presentation. Each slide corresponds to a \textit{stage} within a \textit{scene}, with the data graphic rendering in a full-screen manner, and the corresponding text (or a summary of the text) displayed above it. A \textit{presenter view} is also generated, which contains the relevant controls for the currently displayed slide; a WebRTC connection is established between the presenter's web browser and the browser of each audience member, allowing viewers to see the data graphics update interactively as the presenter manipulates the controls.

\begin{figure*}
  \centering
  % the following command controls the width of the embedded PS file
  % (relative to the width of the current column)
  \includegraphics[width=\textwidth]{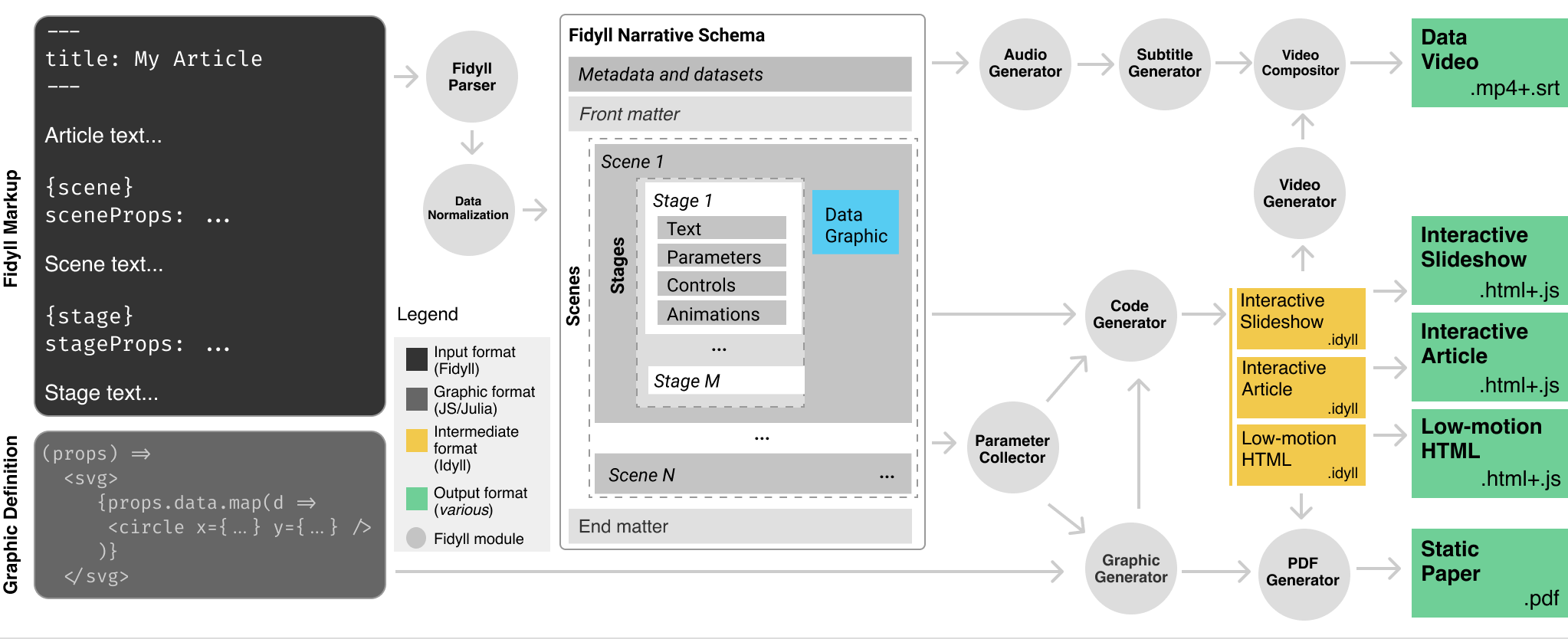}
  % replacing the above command with the one below will explicitly set
  % the bounding box of the PS figure to the rectangle (xl,yl),(xh,yh).
  % It will also prevent LaTeX from reading the PS file to determine
  % the bounding box (i.e., it will speed up the compilation process)
  % \includegraphics[width=.95\linewidth, bb=39 696 126 756]{sampleFig}
  %
%   \parbox[t]{.9\columnwidth}{\relax
%           For all figures please keep in mind that you \textbf{must not}
%           use images with transparent background! 
%           }
  %
  \caption{\label{fig:architecture}
           \fidyll's software architecture. Authors write narrative text in a markup format and provide parameterized graphics. The markup is parsed and normalized  into our schema, which then is used to collect parameter values, generate code, audio, and graphics, and ultimately produce various formats which can be published on the web or elsewhere.}
\end{figure*}

\textbf{Data Video.} The data video target utilizes the same layout as the interactive slideshow. An MPEG-4 video is generated by recording the slideshow played from beginning to end; voiceover is generated by converting the text for each slide to audio via a text-to-speech API and subtitles are generated using the same text in the SRT file format~\cite{brain_subrip_nodate}. This output target does not support interactive control but does include all animations. 

% The slideshow is rendered using a programatically controlled web browser, input events are synthesized to advance through slides, and a screen recording is captured. The text for each slide is converted into audio via a text-to-speech API, and the slide transitions are timed according to the length of the resulting audio recordings. The text is also combined with the audio recording duration information to generate subtitles . The audio recordings are then combined with the video recording to produce the final data video. 

\subsection{Syntax \& Implementation}

\fidyll~is implemented as a Node.js package which can be installed and used via the command line on Linux, MacOS, and Windows. The markup that authors write is based on ArchieML, 
% ~\footnote{http://archieml.org/}, 
a markup language designed to be human-readable and easy for writers and editors to use. ArchieML is used by both technical and non-technical staff across a number of newsrooms, including at the New York Times and Washington Post, and so we believe that our variant could be similarly used by authors without a strong programming background. Example \fidyll markup is shown below:

\begin{verbatim}
---
title: My Great Article
authors: Matthew Conlen, Jeffrey Heer
datasets: 
  penguins: penguins.csv
---

Some introductory text...

{scene} 
graphic: ExampleDataGraphic
parameters: 
  booleanVariable: false
  continuousVariable: 0

This text is shown near the ExampleDataGraphic with both variables in their initial state...

{stage}
parameters: 
  booleanVariable: true

The boolean variable is now true, the ExampleDataGraphic has updated to reflect this.
      
{stage}
animations:
  continuousVariable: 
    start: 0
    end: 1
    duration: 750
controls:
  continuousVariable:
    range: [0, 5, 0.1]

The boolean variable is true and the continuous variable animates from zero to one. Readers can manipulate a range slider to modify the continuous variable after the animation has played, or click a play button to watch the animation play again.
\end{verbatim}

 \textbf{Parsing \& Normalization.} We built a custom parser to handle the new markup format. It reads the YAML-style markup associated with each stage of the article and builds a JSON data structure to represent the schema shown in Figure~\ref{fig:architecture}. Because the parameters that authors provide at each stage may be under-specified (authors only need to specify the parameters that \textit{change}), a normalization step occurs after parsing to ensure that the parameter set associated with each stage of each scene fully specifies all relevant parameters.

 \textbf{Parameter Collection.} Once the data has been normalized into the expected schema, the parameter collector walks through the scenes and stages to identify all of the possible configurations of parameters that could occur while readers progress through an article. These configurations include not only the parameter values specified at each scene and stage, but also the range of parameter values that could be reached by manipulating interactive controls. This set of possible parameter configurations is used to generate any graphics that are created as images on the server-side, and to create static renditions of interactive graphics to be included as appendices in static versions of the article, allowing us to achieve content parity between the static and interactive output formats. 

  \textbf{Graphic Generation.} While many graphics are generated in the browser using web-based libraries like React and D3, we found that many authors also like to create graphics using server-side scripts written in languages like Python, Julia, or R. The graphic generator calls these scripts with all relevant possible parameter values and generates images which can then be embedded in the interactive article. The image filenames encode the parameter values so that interactivity can be maintained through control manipulation (e.g. as a reader manipulates a slider, the slider's value is used to update the path to the image). The graphic generator is also responsible for making static image versions of interactive graphics that ultimately get included as part of an appendix in a static PDF.

 \textbf{Code Generation.} 
 The bulk of the cross-format logic occurs in the code generator, which is responsible for taking in a \fidyll~narrative and producing \idyll markup as output. For each output target, the code generator will iterate through each scene and scaffold the necessary code to generate the relevant layout (e.g. a \textit{Scroller}). \fidyll serializes reactive variables corresponding to parameters (each scene defines its own variable scope) and input widgets and animation playback buttons where appropriate. \fidyll also defines parameter updates via event handlers; the \textit{onEnterView} event, for example, is used to update parameters when a new stage scrolls into view. Much of the code generation logic can be reused across formats, although each has their own unique way of structuring how the content appears on the page. In the static layout, multiple graphical components are embedded in the article, each with their own fixed set of parameters, rather than a single component being embedded which utilizes updating parameters. The code generator ultimately produces an abstract syntax tree representing an \idyll program which is then serialized as markup corresponding to interactive slideshows, scrolling articles, and low-motion articles. This markup is then transformed by the \idyll compiler into HTML and JavaScript that runs in web browsers. It is also used as input to the PDF and Video generators in order to produce static formats. Examples of the compiled \idyll markup are available as part of the supplementary material.

\begin{figure*}[t!]
  \centering
  % the following command controls the width of the embedded PS file
  % (relative to the width of the current column)
  \includegraphics[width=\textwidth]{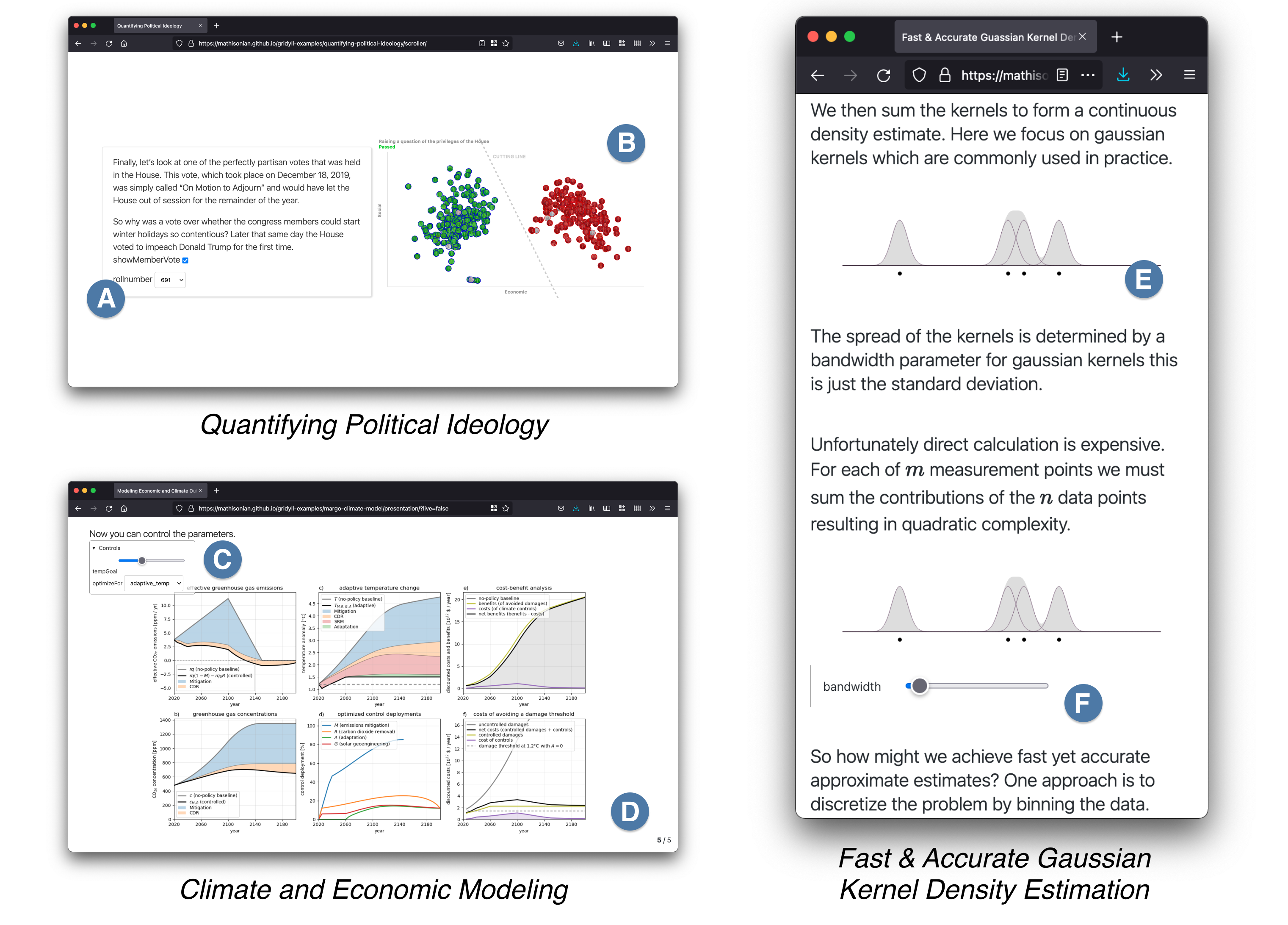}
  % replacing the above command with the one below will explicitly set
  % the bounding box of the PS figure to the rectangle (xl,yl),(xh,yh).
  % It will also prevent LaTeX from reading the PS file to determine
  % the bounding box (i.e., it will speed up the compilation process)
  % \includegraphics[width=.95\linewidth, bb=39 696 126 756]{sampleFig}
  %
%   \parbox[t]{.9\columnwidth}{\relax
%           For all figures please keep in mind that you \textbf{must not}
%           use images with transparent background! 
%           }
  %
  \caption{\label{fig:case-studies}
          We used \fidyll to create three interactive articles. In the interactive article version of \textit{Quantifying Political Ideology}, sections of text scroll on the left half of the screen (A) while graphics remain fixed in place on the right (B); graphics update as the reader scrolls and respond to interactions with controls embedded in the text. The presentation version of \textit{Climate and Economic Modeling} provides high-level text summaries of visualizations and embedded controls (C), which can be used to manipulate full-screen graphics (D). In the low-motion HTML version of \textit{Fast \& Accurate Gaussian Kernel Density Estimation}, graphics are repeatedly embedded in a single text column with various parameters (E); readers can still manipulate controls to further explore the parameter space (F).} 
\end{figure*}

 \textbf{Video Generation.} The video generator takes as input the ~\fidyll~narrative and runs the text through a text-to-speech API to create a series of audio snippets, one corresponding to each slide in the interactive slideshow. We use the Text-to-Speech service provided by Google Cloud. Once the audio is created, the interactive slideshow is recorded via a headless web browser~\cite{puppeteer}. The duration of each audio snippet is use to determine the length of each slide in the interactive slideshow as well as the timings for the generated subtitles. The audio clips and screen recording are composited together to produce an MP4 video and an SRT subtitle file is serialized.

\textbf{PDF Generation.} The PDF generator leverages the low-motion interactive web article as well as the document converter Pandoc~\cite{macfarlane2013pandoc}. Each graphic in the low-motion article is converted into an image and an HTML file is serialized with the interactive graphics replaced by static images. 
% The HTML file is also augmented to include additional metadata that helps Pandoc produce the final PDF. 
The HTML file is processed by the Pandoc converter, which produces a PDF using \LaTeX.

\section{Evaluation}

We consider this work to be a form of user interface systems research and turn to Olsen's criteria for evaluation~\cite{olsen2007evaluating}. In Sections \ref{sec:introduction} \& \ref{sec:bg} we showed the \textit{importance} of the work and in Sections \ref{sec:interviews} \& \ref{sec:motivating-scenarios} we identified specific \textit{situations, tasks and users}. In this section we show the \textit{generality} of our work through a set of three case studies, and show how it \textit{reduces solution viscosity} compared to existing tools by affording authors \textit{expressive leverage}: \fidyll significantly reduces the amount of non-narrative markup that authors need to write to generate cross-format interactive articles.

\subsection{Case Studies}
\label{sec:evaluation:case-studies}

We used~\fidyll~to develop three cross-platform interactive articles (Figure~\ref{fig:case-studies}). These articles were designed so that each would map directly on to one of the motivating scenarios provided in \S\ref{sec:motivating-scenarios}. The articles cover a range of topics (spatial models of politics; an economic-climate model; an optimization for kernel density estimation) using a variety of commonly used narrative visualization techniques such as a \textit{martini glass} structure and \textit{drill-down} stories~\cite{segel2010narrative}. Each of the articles was developed with the intention that it would serve as a useful educational artifact separate from the publication of this paper, which helps ensure that they provide ``authentic and realistic motivation'' for this evaluation~\cite{2019-idyll-analytics}. The articles are available online at \url{https://idyll-lang.github.io/fidyll-examples/} and a video demo is provided as supplementary material.
% \edit{Say something about when/where they were published?}

\subsubsection{\textit{Quantifying Political Ideology}}

In \textit{Quantifying Political Ideology} we provide an interactive explanation of the NOMINATE family of spatial models of rollcall voting for the U.S. legislative branch~\cite{poole1985spatial, poole2005spatial}. DW-NOMINATE scores are frequently cited by journalists who use them as a tool for interpreting political behaviors. The article is written from the perspective of a data journalist who wants to provide their audience with a high-level and intuitive explanation of a topic that is typically presented in a highly technical way.  The article consists of three scenes: an \textit{introduction} which explains the basic properties of spatial voting models and provides a high-level motivation for why they are built; \textit{real world examples}, in which the voting behavior and ideological positions of members of the 116th U.S. Congress is explored (readers can interactively select votes which they are interested in learning more about); and a \textit{technical explainer} of the inner-workings of the optimization algorithm used to derive the ideological NOMINATE scores.

\subsubsection{\textit{MARGO Climate Modeling}}

\textit{MARGO Climate Modeling} was designed to match the motivating scenario MS2, where a non-profit organization uses open datasets and scientific models in order to advocate for certain policy actions. In this article we utilize the MARGO model~\cite{drake2020multi} which allows users to specify constraints (e.g. \textit{``keep warming below two degrees celcius''}) and will provide an optimized path for adhering to those constraints through the use of climate controls including mitigation of greenhouse gas emissions, adaptation to climate impacts, removal of carbon dioxide, and geoengineering. The article is a straightforward application of the martini glass narrative structure in which an author-driven narrative is first presented to the reader, walking them through the a series of important points pertaining to the MARGO model, and then allowing them to explore freely within a predefined space of model parameters.  Because the model runs in Julia,~\fidyll determined the multidimensional space of possible parameter values that a reader could reach at any point in the article, and for each possible configurations, ran the model ahead of publication to generate corresponding graphics. In the final version there were 22 possible parameter configurations, which took just under an hour to generate on a 2018 MacBook Air with a 1.6 GHz dual-core processor and 16 GB of ram.

\begin{table*}\centering
\small
\begin{tabular}{@{}p{1.5cm}rrccrrrrr@{}}\toprule Example & Scenes & Stages & Target & Markup & 
\multicolumn{1}{p{1cm}}{\centering Narrative\\LOC} &
\multicolumn{1}{p{2cm}}{\centering Non-narrative\\LOC} & \multicolumn{1}{p{1cm}}{\centering Total\\LOC} & 
\% Narrative & \multicolumn{1}{p{2cm}}{\centering \% Reduction\\(Non-narrative)} \\ 
\midrule

Fast KDE & 18 & 54
& & fidyll & 109 & 320 & 429 & 25.41\% &  
\\
& & & static & idyll & 109 & 1518 & 1627 & 6.70\% & 78.92\%
\\
& & & slideshow & idyll & 109 & 1083 & 1192 & 9.14\% & 70.45\%
\\
& & & scroller & idyll  & 109 & 736 & 845 & 12.90\% & 56.52\%
\\& & & \textit{combined} & idyll & 109 & 3337 & 3446 & 3.16\% & \textbf{90.41}\%
\\
\\
MARGO & 2 & 14
& & fidyll & 33 & 95 & 128 & 25.78\% &  
\\
& & & static & idyll & 33 & 179 & 212 & 15.57\% & 46.93\%
\\
& & & slideshow & idyll & 33 & 219 & 252 & 13.10\% & 56.62\%
\\
& & & scroller & idyll & 33 & 142 & 175 & 18.86\% & 33.10\%
\\& & & \textit{combined} & idyll & 33 & 540 & 573 & 5.76\% & \textbf{82.41}\%
\\
\\
QPI & 2 & 13
& & fidyll & 57 & 90 & 147 & 38.78\% &  
\\
& & & static & idyll & 57 & 495 & 552 & 10.33\% & 81.82\%
\\
& & & slideshow & idyll  & 57 & 256 & 313 & 18.21\% & 64.84\%
\\
& & & scroller & idyll  & 57 & 190 & 247 & 23.08\% & 47.37\%
\\& & & \textit{combined} & idyll & 57 & 941 & 998 & 5.71\% & \textbf{90.44}\%
\\
\bottomrule\end{tabular}\caption{\fidyll~reduces solution viscosity by providing authors with expressive leverage. Authors using \fidyll~write considerably less non-narrative code than with existing markup languages, freeing them to focus on the narrative aspects of their data story.}\label{fig:loctable}\end{table*}

\subsubsection{\textit{Fast Kernel Density Estimation}}

The \textit{Fast Kernel Density Estimation} article was developed to represent MS3, the scientific publishing scenario, and was adapted from a presentation given by the final author of this paper at the 2021 \textit{IEEE Visualization Conference}~\cite{heerfast}. The original presentation was created by having the presenter give the talk into his webcam while screen recording software captured slides produced in Keynote. The slides contained multiple graphics which were developed in JavaScript using D3~\cite{bostock2011d3} and utilizing both Observable~\cite{observable} and a standard (non-notebook) web development environment. The graphics were interactive, but for the purposes of the presentation multiple video recordings were made with screen recording software and then added to the slides. We were able to adopt these graphics for use within~\fidyll~and reproduce the content of presentation as a cross-platform explorable explanation, including an interactive slideshow in which the graphics can be manipulated in realtime by the presenter. To do so we needed to make small modifications to the code of the graphics (approx. 10 lines of code changes in each graphic) so that they adhered to our component API and can adapt to different screen sizes (the graphics are displayed at a different resolutions in different formats, e.g. in the static article they are shown at the width of the text column; in the interactive slideshow they are shown at full screen width).

\subsection{Expressive Leverage}
 
To assess the extent to which \fidyll~assists in facilitating the rapid creation of cross-format interactive articles we analyzed the source code of the three case study articles, including both the \fidyll~markup that authors produced as well as the intermediate \idyll files which get created and ultimately compiled into HTML and JavaScript. We compare the \fidyll markup to the intermediate \idyll markup rather than the final HTML and JavaScript because \idyll represents a more realistic baseline in terms of what authors currently use for the task of authoring interactive articles. We computed relevant counts and ratios of the lines of code (LOC) associated with each case study and with the three intermediate \idyll formats which ultimately generate the five final artifacts. Table~\ref{fig:loctable} shows the results of this analysis. To make the comparison valid across the articles and formats we break each sentence of narrative text onto its own line and ignore any whitespace. The \idyll markup that is generated is idiomatic markup that is produced by calling the \textit{idyll-ast} package's \textit{toMarkup(ast)} function. We consider a narrative line of code to be any line in which the majority of the markup is rendered directly on-screen as narrative text; the \% reduction metric is defined as one minus the lines of non-narrative \fidyll code divided by the lines of non-narrative \idyll code.

 We found that \fidyll reduced the amount of non-narrative code by 82–90\% compared to writing the \idyll markup for each of the three \idyll formats directly. Compared to writing the \idyll markup, for a just a single format \fidyll still reduced the amount of code written by 33–82\% (mean 60\%; median 57\%).  \fidyll also improved the ratio of narrative to non-narrative code that authors had to write: the \fidyll markup consisted of 30\% narrative code on average, while this number was 15\% on average for a single format drafted in \idyll and 5\% for the combined \idyll markup, indicating that the authors would have had to write about 20 lines of non-narrative markup for every sentence in the data story.

\section{Discussion}

We discuss the limitations of our system and study and highlight points of interest to the research community.

\textbf{Limitations of this work}. While we grounded the situations, tasks, and users of our system through formative expert interviews and created a set of realistic case studies, we haven't directly observed our system in the hands of end users. Because our markup language is a dialect of ArchieML, a widely used markup language in newsrooms, we don't anticipate users having an issue learning the syntax. In the future we would like to observe the strengths and weaknesses of \fidyll through observation of real world use. While \fidyll can express a wide range of common narrative visualization design devices, there are designs that it does not support. For example, the tool expects that readers will progress through content in linear fashion and does not support non-linear \textit{choose-your-own-adventure}-style text. The tool only supports a subset of possible formats and doesn't yet support highly-visual genres of narrative visualization such as data comics or infographics. 

\textbf{The Role of Automation in Data Storytelling}. While our system drastically reduces the LOC needed to produce multi-format data stories, there are challenges to the automated approach~\cite{heer2019agency}. For example, the audio narration (generated by a text-to-speech model) frequently mispronounces technical words (e.g. ``Gaussian'') and in practice we expect that authors will prefer to add voiceover from a human narrator. We attempted to automate other tasks, such as using a language model~\cite{brown2020language} to summarize text from long form articles to shorter snippets that would be appropriate in formats like a slideshow. We achieved poor results; however, this may be useful to develop further in the future. Our system gives authors access to all of the generated assets so that they can make additional manual changes if necessary. 
% This aligns with the current practice that we identified during our interview in which authors frequentlly 

% \edit{Add callback to interviews and how people make manual tweaks.}

\textbf{Parity of Content Across Formats}. While there are real educational benefits to well-designed interactive content~\cite{hohman2020communicating}, there are still situations where static or non-interactive animated content will be preferred. In these cases we believe it is important to still give readers access to the same \textit{information content}~\cite{tversky2002animation}. For this reason we ensure that static versions always included additional still images of graphics, representing all of the relevant states that could be reached in the interactive version. 
% \edit{Add inspiration / future work, what should future versions of the system do?}
 
\textbf{Accessibility of Interactive Articles}. \fidyll~supports authors in producing accessible documents: the static and low-motion formats are preferable for readers who experience motion sickness; the audio and subtitles generated for the video format support readers with low vision. However, there are still major challenges in adapting interactive graphics for use with assistive devices like screen readers~\cite{kim2021accessible}; care needs to be taken to generate markup that works well with these devices, and the system should create descriptions of the on-screen graphics for low-vision readers.  
% \edit{What else?}
 
\textbf{System Expressiveness and Design Guidance}. While our system is strictly less expressive than \idyll, which our high-level markup compiles to, we believe that this trade-off is beneficial for many authors because of the expressive leverage that it provides, and because it provides much more structure in the resulting designs, nudging authors to use common, well-established formats. Many authors of interactive articles are not experts in design and likely benefit from the system making low-level design decisions.

% \edit{Do we have a citation for this?}

\section{Conclusion}

We presented \fidyll, a cross-format compiler for data stories and explorable explanations. Our system was informed by interviews with expert users and realistic motivating scenarios. The scenarios were operationalized in three case studies that we analyzed to show the expressive leverage of our system, reducing the amount of non-narrative code that needs to be written by 88\% on average. 

%-------------------------------------------------------------------------
% bibtex
\bibliographystyle{eg-alpha-doi}
\bibliography{egbibsample}

% biblatex with biber
% \printbibliography                

\end{document}